\documentclass[twocolumn,showpacs]{revtex4}

\usepackage{graphicx}
\usepackage{indentfirst}
\usepackage{amsfonts}
\usepackage{amsmath}
\usepackage{amssymb}
\usepackage[T1]{fontenc}
\newcommand{\ket}[1]{\left| #1 \right\rangle}
\newcommand{\bra}[1]{\left\langle #1 \right|}

\newcommand{\eg}{{\textit{e.g.}}, }
\newcommand{\ie}{{\textit{i.e.}}, }

\begin{document}

\title{Asymptotic Entanglement Dynamics and Geometry of Quantum States}

\author{R.~C.~Drumond$^{\dagger}$}
\author{M.~O.~Terra~Cunha$^{\ddagger}$}

\affiliation{$^{\dagger}$Departamento de F\'{\i}sica, Instituto de
Ci\^{e}ncias Exatas, Universidade Federal de Minas Gerais, CP 702,
CEP 30123-970, Belo Horizonte, Minas Gerais, Brazil}
\affiliation{$^{\ddagger}$Departamento de Matem\'{a}tica, Instituto
de Ci\^{e}ncias Exatas, Universidade Federal de Minas Gerais, CP
702, CEP 30123-970, Belo Horizonte, Minas Gerais, Brazil}

\begin{abstract}
A given dynamics for a composite quantum system can exhibit
several distinct properties for the asymptotic entanglement
behavior, like entanglement sudden death, asymptotic death of
entanglement, sudden birth of entanglement, etc. A classification of
the possible situations was given in [M.~O.~Terra Cunha, {\emph{New J. Phys}}
{\bf{9}}, 237 (2007)] but for some classes there were no known examples. In this work we
give a better classification for the possibile relaxing dynamics at the
light of the geometry of their set of asymptotic states and give
explicit examples for all the classes.
Although the classification is completely general, in the search of examples it is
sufficient to use two qubits with dynamics given by
differential equations in Lindblad form (some of them
non-autonomous). We also investigate, in each case, the probabilities to
find each possible behavior for random initial states.
\end{abstract}


\maketitle

\section{Introduction}

Entanglement is a fundamental property of composite quantum systems,
first noted by Schr\"{o}dinger \cite{schroedinger}. The best
knowledge of the whole of a composite quantum system may not include
complete knowledge of its parts. It has strong conceptual
implications on physics, since it is a property that has no
classical analog, so we are forced to change significantly our
perspective of Nature. Such peculiar character allows it to be
considered as a fundamental resource for some non-classical tasks
as teleportation of a quantum state \cite{teleporte}, quantum
computation \cite{quantumcomp}, quantum cryptography \cite{ekert},
etc \footnote{Entanglement is not necessary for quantum key distribution,
however it is used in the best known proof of security of such protocols \cite{renner}}.
Once entanglement is considered a resource it seems natural
to quantify it \cite{entanquant}. In all the applications named
above, it is necessary to optimize the amount of entanglement in a
suitable composite quantum system to best execute the desired task.

Real quantum systems always interact with its environment, irrespectively of
the efforts to protect
it. This interaction will, in general, create
some entanglement between the quantum system and the environment,
and this entanglement will, somewhat ironically, spoil the
entanglement between the parts of the ``useful'' system (for
bipartite systems, this affirmation has a precise meaning provided by the
monogamy of entanglement theorem \cite{monogamy}).

While in most of the models used to describe quantum open systems
the coherences of a state decays asymptotically to zero, it was
recently recognized that entanglement may ``die'' at finite time
\cite{primeiramorte}, a phenomenon called entanglement sudden death
\cite{eberlyesd}. This phenomenon has called some attention,
specially connected to the difficulty of keeping entanglement alive
for its uses as a resource. Some interesting generalizations were
studied \cite{gen}, and some experiments were proposed
\cite{proposta} and realized \cite{experimentodavidov}. This
phenomena, though, has a simple explanation if one looks at the
geometry of quantum states \cite{terra}. Namely, while the set of
``decohered'' states always have zero volume inside the set of all
possible quantum states, the set of separable states has not only a
positive volume but also non-empty interior \cite{volume}
when the global system have a finite dimensional Hilbert space.

The geometrical approach to the problem allows one to classify the
dynamics of a quantum system according to the geometry of its
asymptotic states (if the dynamics implies them) relative to the set
of separable states \cite{terra}. In the cited paper
some classes were exemplified, but to that time it was not clear whether
all a priori
possible situations could be found.

In this paper we review the geometric classification of entanglement dynamics
and provide explicit examples to all
a priori possible situations. All examples are given in
the two-qubit Lindblad differential equations context, with some
cases using non-autonomous equations (exactly those in the classes
for which examples were not previously known). We also introduce a new
analysis of how often each specific behavior occur for a given
dynamics, in the light of probability theory applied to the set of
initial states \cite{viviescas}.

\section{The Geometry of Entanglement Sudden Death: General
Picture}\label{eventos}

\begin{figure}[htp]
  \centering
  \includegraphics[width=4.5 cm]{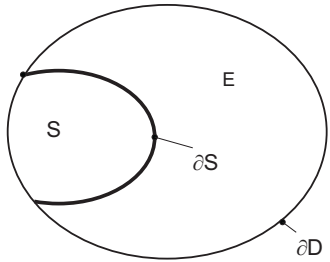}\\
  \caption{Diagram of the set of entangled states. }\label{figura1}
\end{figure}

What can we say about the geometry of entanglement, or the geometry
of the set of separable states, for general multipartite systems?
First of all, that the set of separable states is closed, convex and
with non-empty interior (we shall assume finite dimensional Hilbert
spaces throughout the paper). Its complement relative to the set of
quantum states also has non-empty interior and is certainly
non-convex. Actually, in general, its complement is much larger,
i.e., it has greater volume (if one consider the Hilbert-Schmidt
metric, for instance). An extremely oversimplified illustration of
this situation is given in Fig. \ref{figura1}. We call here $D$ the
set of all quantum states and we are going to consider it immersed
in the set $\mathcal{A}$ of Hermitian matrices of unity trace; $S$
the subset composed by the separable states, $\partial S$ and
$\partial D$ their boundaries relative to $D$ and $\mathcal{A}$,
respectively; $E=D-S$ the set of entangled states. The boundary of
the set of quantum states is composed by all that states which have
at least one zero eigenvalue so, in particular, it contains all the
pure states. Note that there are both entangled and separable pure
states in $\partial D$. Actually, more than that, the ``area'' of
the separable states inside $\partial D$ is non-zero
\cite{artigojgp}.

Let us consider a dynamics with a non-trivial stationary
set, $St$. By stationary set we mean that for every initial
state $\rho$ and open set $V\supseteq St$ we have that $\rho(t)$
(the state at time $t$) belongs to $V$ for all $t$ sufficiently
large. Of course, if some dynamics accepts a set of stationary
states, this set will be the smallest stationary set of the
dynamics. Anyway, from the simple picture given in Fig.~\ref{figura1}, and
considering the location of $St$ in it, we may distinguish three
possibilities which have consequences to the asymptotic dynamics of
entanglement: i) $St\subset\text{ Int}(S)$ implies that every
initial entangled state will lose all of its entanglement at finite
time (sudden death of entanglement); ii) if $St\cap
\partial S\neq\emptyset$, than, only with that information, many
situations can occur: asymptotic or sudden death of entanglement and
non-zero asymptotic entanglement; iii) if $St\subset E$, every
initial state exhibit some entanglement asymptotically.

The complete classification must yet consider that the stationary
set, $St$, can consist of a single state (\eg thermal equilibrium state)
or by a non-trivial set (\eg for phase reservoirs). In this sense, each
situation above gives rise to two cases, in a total of six classes.

Note that, in cases ii) and iii), if we start with a separable state
it is possible in the first one and certain in the second, that
entanglement will be created, a situation which may be called sudden
birth of entanglement \cite{terra}. It is good to stress that, since
the only information we have about the dynamics is some partial
information about a stationary set, anything may happen with the
entanglement for short times: it may die, resurrect, oscillate, etc.
It is also important to mention that such analysis does not depend
on the specific entanglement quantifier used to follow the dynamics,
only the assumption that it is continuous and strictly positive on
entangled states.

Given a dynamics that fits in case ii) one can in general find
examples of initial states whose entanglement die asymptotically or
suddenly \cite{proposta}. An interesting way to have a global view
of the properties of this dynamics on this respect is through the
question: if one pick a random initial state, what is the most
probable situation, asymptotic or sudden death? That is, if the
dynamics can exhibit both of these properties, what is the most
typical one? To answer this question one must formulate
it properly. Fixed a dynamics for a composite system with state
space $D$ with a suitable probability measure $P$ on it and a
continuous entanglement quantifier $e:D\rightarrow \mathbb{R}_{+}$,
with $e(E)\subset (0,\infty)$, we define the following events
(subsets of $D$, in the language of probability theory) whose
probabilities may be of interest:
\begin{itemize}
       \item States that exhibit sudden death of entanglement: $SDE=\{\rho\in \mathcal{D}_{N}|\exists t_{0}, t_{1} \text{ such that } E(\rho(t_{0}))>0 \text{ and } E(\rho(t))=0 \text{
       for all }
       t>t_{1}\}$;
       \item States that exhibit asymptotic death of entanglement: $ADE=\{\rho\in \mathcal{D}_{N}|\exists (t_{n})_{n=1}^{\infty}, t_{n}\rightarrow\infty, \text{ such that } E(\rho(t_{n}))>0 \text{ and }
       \text{lim}_{t\rightarrow\infty}E(\rho(t))=0\}$;
   \end{itemize}
where $\rho(t)$ denotes the time $t$ evolution of initial state $\rho$ according to the dynamics.
Note that these definitions do not coincide strictly with the
common sense of such notions since in general one only looks for
initial states that already have some entanglement, which is not
necessary here: an initial separable state can, in principle,
acquire some entanglement that will subsequently die (suddenly or
asymptotically). The strict notion would be given by the events:
     \begin{itemize}
       \item $SDE'=SDE\cap E$;
       \item $ADE'=ADE\cap E$.
     \end{itemize}
If the dynamics exhibit asymptotic entangled states, one can also
look to the events:
   \begin{itemize}
       \item The states exhibit entanglement asymptotically: $AE=\{\rho\in \mathcal{D}_{N}|\exists t_{0}, c>0 \text{ where } E(\rho(t))>c \text{ for all }t>t_{0}\}$;
       \item An initially separable state acquire entanglement asymptotically (sudden birth of entanglement): $SBE=\{\rho\in \mathcal{D}_{N}|E(\rho)=0 \text{ and } \exists t_{0}, c>0 \text{ where }E(\rho(t))>c \text{ for all } t>t_{0}\}$;
  \end{itemize}
(note that $SBE=AE\cap S$).

Instead of choosing a specific probability measure to deal with, our
results will only require that it is non-singular, i.e., sets
contained in sub-manifolds of $D$ with dimensions strictly smaller
than the dimension of $D$ have zero probability. The problem of
computing the probability (or volume) of the event (set) $S$ exactly
is still an open issue for the most natural probability measures.
Though, several bounds and estimates exist for several probability
measures and events \cite{artigojgp, geometria}.

\section{Explicit Examples}

Given the general picture we may look now to some concrete examples
where most of them, as we will see, are very natural and
experimentally feasible. The simplest type of dynamics, namely
one that is convex-linear, Markovian and completely positive, will
suffice to provide rich examples. It will be sufficient to work with
the simplest composite system, two qubits, in order to exhibit
examples for all classes of dynamics. Considering that we are
dealing with a system with finite Hilbert space, we may apply
Lindblad theorem and describe the map by an ordinary, linear, first
order differential equation with the form given by \cite{lindblad}:
\begin{subequations}
\begin{equation}
 \frac{d\rho}{dt}=\mathcal{L}[\rho]=-\frac{i}{\hbar}[H,\rho]+\mathcal{D}[\rho],
\end{equation}
where
\begin{equation}
\mathcal{D}[\rho]=\sum_{j} \gamma_{j}(2A_{j}\rho
A^{\dagger}_{j}-A^{\dagger}_{j}A_{j}\rho-\rho A^{\dagger}_{j}A_{j}),
\end{equation}
\end{subequations}
$\gamma_{j}$ are real constant numbers, $A_{j}$ general linear
operators, $H$ a Hermitian operator. This type of dynamics have the
advantage that it is simple to find its asymptotic states: in
general one just have to look to the kernel of the ``superoperator''
$\mathcal{L}$, which is a linear operator that can be understood to
be defined over the set of $4\times 4$ complex matrices or the
subset of Hermitian matrices, a real vector space.
Of course, since the set will be given by the kernel of
linear map, it is always given by the intersection of a subspace of
Hermitian matrices (the kernel of $\mathcal{L}$) with the set of mixed states.
It is courious to note that to
find the two missed examples in Ref.~\cite{terra} we had to allow for non-autonomous
Lindblad equations, that is, equations with the same form but with
parameters $\gamma_{j}$ varying in time. For this type of dynamics
the set of stationary states do not need to be the intersection of a
subspace with the set of quantum states.

\subsection*{Localizing a two qubit state in the set of all states}

The set of all quantum states for a composite system can be divided
geometrically according to the dichotomy $\left\{\text{Int}D,\partial
 D \right\}$ and the trichotomy $\left\{\text{Int}S, \partial S,
 E\right\}$. Dealing with the special case of two
qubits has the advantage that one can easily infer the location of a
state according to this subdivision with the help of
$\text{Det}\rho$ and $\text{Det}\rho ^{\Gamma}$, the determinants of
the state and of its partial transpose. Both of these functions are
continuous in all natural metrics, Hilbert-Schmidt, etc, i.e, we
know that small perturbations of a state in a given metric implies
small perturbations of the values of both quantities. So if,
{\it{e.g.}}, both of them are positive for a given state, we can
find a neighborhood of that state where these quantities remain with
the same sign. Then, the determinant of the operator tell us if it
is in the interior or in the border of $D$ (if it is grater than or
equal to zero, respectively). The determinant of the partial
transpose, on the other hand, gives us complete information about
its entanglement because it is known \cite{dettransp} that the state
is entangled iff the determinant is strictly negative. Thus, this
determinant tells us if the state is in the interior of the set of
separable states (if it is grater than zero), in the border
$\partial S$ (if it is equal to zero) or inside the set of the
entangled states (if it is strictly negative). At last, if $\rho$ is
a given state with $d=\text{Det}\rho$ and
$d^{\Gamma}=\text{Det}\rho^{\Gamma}$, we can have:
\begin{itemize}
\item[i)] $d>0$ and $d^{\Gamma}>0$: the state is in the interior of $D$ and in the interior of $S$ relative to $D$, i.e., belongs to $S-\partial S$ ({\it{e.g.}}, the completely mixed state,
$\rho_{mix}=I/4$);
\item[ii)] $d>0$ and $d^{\Gamma}=0$: the state is in the interior of $D$ and in $\partial S$ ({\it{e.g.}}, the state
$\frac23\rho_{mix}+\frac13\rho_{singlet}$, where $\rho_{singlet}$ refers to the state in Eq.~\eqref{eq:2} with $a=0$, $b=-c=1/2$);
\item[iii)] $d>0$ and $d^{\Gamma}<0$: the state is in the interior of $D$ and belongs to $E$ ({\it{e.g.}}, the Werner states \cite{werner} $p\rho_{mix}+(1-p)\rho_{singlet}$,
for $0<p<2/3$);
\item[iv)] $d=0$ and $d^{\Gamma}>0$: the state is in the border of $D$ and in $S-\partial S$ (recording that we defined $\partial S$ as the boundary relative to $D$, while $\partial D$ is relative to $\mathcal{A}$). For instance, if $a>b>0$, $2a+2b=1$, and
$|c|=b$:
\begin{eqnarray}\label{eq:2}
\rho= \left( \begin{array}{cccc}
a & 0 & 0 & 0\\
0 & b & c & 0\\
0 & c & b & 0\\
0 & 0 & 0 & a
\end{array} \right);
\end{eqnarray}
\item[v)] $d=0$ and $d^{\Gamma}=0$ the state is in $\partial D\cap \partial S$ ({\it{e.g.}}, a separable pure state);
\item[vi)] $d=0$ and $d^{\Gamma}<0$ the state is in the border of $D$
and belongs to $E$ ({\it{e.g.}}, $\rho_{singlet}$).
\end{itemize}

With all these tools in hand, we can go to the examples.

\subsection*{Case 1a): One asymptotic state in Int$(S)$}

Perhaps the most natural example of this situation is the case where
both qubits, which we shall call $A$ and $B$, are spatially well
separated two level atoms, interacting with thermal fields. The
separation between them implies that the thermal reservoirs are
independent. The Lindblad equation that describes this dynamics is
given by:
\begin{subequations}\label{thermal}
\begin{equation}
\frac{d\rho}{dt}=\frac{i}{\hbar}[H_{A}+H_{B},
\rho]+\mathcal{D}_{A}\otimes I[\rho]+I\otimes\mathcal{D}_{B}[\rho],
\end{equation}
where
\begin{align}\label{decaydissipator}
\mathcal{D}_{i}[\rho]=\gamma_i(2\sigma_{+,i}\rho\sigma_{-,i}-\sigma_{-,i}\sigma_{+,i}\rho-\rho
\sigma_{-,i}\sigma_{+,i})+ \nonumber \\
\gamma'_i(2\sigma_{-,i}\rho\sigma_{+,i}-\sigma_{+,i}\sigma_{-,i}\rho-\rho
\sigma_{+,i}\sigma_{-,i}),
\end{align}
\end{subequations}
with $\sigma_{\pm,i}$ being the Pauli operators for qubit $i$,
$\displaystyle{H_i = \frac{\hbar\omega_i}2 \sigma_{z,i}}$ the Hamiltonian for qubit $i$ and $\gamma_i, \gamma'_i$
are non-negative constants (related to the average photon number in
the field, the atoms polarization, their coupling to the environment, etc.).

It is easy to show that the system will evolve to a product state
with both qubits in their respective Gibbs states, $Z_i^{-1}e^{-\beta H_i}$,
$Z_i = \text{Tr} e^{-\beta H_i}$.
If the temperature is positive, the resulting state is a product state
with a diagonal density matrix  (in the product basis) with every
diagonal entrance being non-zero. We then have that
$\rho_{st}=\rho_{st}^{\Gamma}$ and that
$\text{Det}\rho^{\Gamma}=\text{Det}\rho>0$. As mentioned earlier, if
an initial state have some entanglement it will certainly die at
finite time.

The events defined in sec.~\ref{eventos} are trivial in this case:
$ADE=ADE'=SBE=AE=\emptyset$, and $SDE=SDE'=E$, so $P(SDE)=P(E)$ or,
$P(SDE|E)=1$, that is, the only condition for having entanglement
sudden death is that the initial state is entangled. To calculate
the exact probability of this event is thus as difficult as determining the
volume of the set of separable states \cite{artigojgp}.

\subsection*{Case 1b): Several asymptotic states in Int$S$}

To obtain an equation of motion for the state satisfying this
propriety, namely, being a relaxing dynamics with more than one
asymptotic state but all of them in the interior of $S$, we had to
appeal to a non-autonomous Lindblad equation. A dynamics that
achieve the desired result would be given by a Lindblad equation
with the same form as the one used in the last section, describing
two qubits interacting with independent reservoirs, but now, with
the coupling ``constants'' decaying exponentially. That is, performing
the correspondence $\gamma_i\mapsto \gamma_{i0}\exp{(-\kappa t)}$. The
physical situation corresponding to the equation, although
artificial, is certainly not prohibited: in principle, one can have
a good control of the interaction of the qubits with their reservoir
and turn it off exponentially.

To prove the result, let us write the dynamical equation in the form (in the
interaction picture):
\begin{equation}\label{equacao1}
\frac{d}{dt}\rho(t)=e^{-\kappa t}\mathcal{D}[\rho(t)],
\end{equation}
where $\mathcal{D}$ is the dissipator of the Lindbladian in the last
example. For $\rho(t)$ a solution to this equation, we can
define  $\bar{\rho}(t)=(\rho\circ g)(t)$, where
\begin{displaymath}
g(t)=\int_{0}^{t}e^{-\kappa t'}dt'
\end{displaymath}
is an invertible function. Substituting $\bar{\rho}$ in
Eq.~\eqref{equacao1} we obtain an equation of motion for it:
\begin{equation}\label{equacao2}
\frac{d}{dt}\bar{\rho}(t)=\mathcal{D}[\bar{\rho}(t)].
\end{equation}
That is, $\bar{\rho}$ obeys the same dynamics of two qubits in
independent thermal reservoirs with constant coupling in time, with
known solution. To find the asymptotic set for the dynamics of
Eq.~\eqref{equacao1} is sufficient to note that, since
$\rho(t)=(\bar{\rho}\circ g^{-1})(t)$, then
$\rho(t\rightarrow\infty)=\bar{\rho}(g^{-1}(t\rightarrow\infty))=\bar{\rho}(1/\kappa)$.

Geometrically, the autonomous dynamics given by Eq.~\eqref{equacao2}
deforms continually the set of states $D$ to the point
$\rho_{Gibbs}$ (\ie provides an homotopy between them), while the time
varying version reparametrizes this deformation. The set of
asymptotic states of Eq.~\eqref{equacao1} is then given by this
deformation in the intermediate time $\kappa^{-1}$. Making $\kappa$
small enough, we can assure that the asymptotic set is entirely
contained in Int$S$, since $\rho_{Gibbs}$ belongs to Int$S$, an open
set.

Of course, the events $SDE$, $ADE$, etc., and their respective
probabilities, are exactly the same as in the last example.

We note finally that, although the discussion about entanglement
does not depend if we are dealing with the interaction or
Schr\"{o}dinger pictures (because the correspondence between them is
given by local unitary transformations), the dynamics is not
relaxing in the former. Since the state will be given by
$\rho_{S}(t)=\exp(iHt)\rho(t)\exp(-iHt)$ and
$\lim_{t\rightarrow\infty}\rho(t)$ will not, in general, commute
with the exponentials, the state evolution $\rho_{S}(t)$ will not
converge. Nevertheless, the dynamics will have an asymptotic set in
the general sense discussed in Sec.~\ref{eventos}, namely, although
an initial state does not necessarily converges, one can find open
sets such that the state trajectory will be confined inside them
after a certain instant of time. In this particular example, one
can find such open sets that are entirely contained in $S$.

\subsection*{Case 2a): One asymptotic state in $\partial
S$}

Eqs.~\eqref{thermal} also provides an example where we have only one
stationary state in the border between separable and entangled
states, namely, the case where the qubits are subjected to two
independent thermal reservoirs at \emph{null temperature}. In this
case the stationary state is the pure state
$\rho_{st}=\ket{00}\bra{00}$. Again, it is diagonal in the
computational basis so
$\text{Det}\rho_{st}^{\Gamma}=\text{Det}\rho_{st}=0$. Then, a
neighborhood of this state always contains separable as well as
entangled states. As mentioned in Sec.~\ref{eventos}, in this example,
depending on the initial state, both behaviors can happen:
asymptotic and sudden death of entanglement.
In fact, given an initial state with matrix
elements $\rho_{ij}$ one can shown that the determinant of the
partial transpose of the state in time $t$ will be given by:
\begin{subequations}
\begin{equation}
  \text{Det}\rho^{\Gamma}(t)=e^{-4\kappa
  t}\text{Det}[\rho'+\rho''(t)],
\end{equation}
where
\begin{equation}
\rho'=\left( \begin{array}{cccc}
\rho_{11} & \rho_{12}^{*} & \rho_{13} & \rho_{23}\\
\rho_{12} & \rho_{11}+\rho_{22} & \rho_{14} & \rho_{24}+2\rho_{13}\\
\rho_{13}^{*} & \rho_{14}^{*} & \rho_{11}+\rho_{33} & \rho_{34}^{*}+2\rho_{12}^{*}\\
\rho_{23}^{*} & \rho_{24}^{*}+2\rho_{13}^{*} & \rho_{34}+2\rho_{12}
& 1
\end{array} \right),
\end{equation}
\end{subequations}
and $\rho''(t)$ is a matrix which depends on $\rho$ but where all
elements decay (exponentially) to zero. Hence, as long as
Det$\rho'\neq 0$, the asymptotic sign of
$\text{Det}\rho^{\Gamma}(t)$ will be given by the sign of the
determinant of $\rho'$. By assuming non-singular probability measure
in the set of quantum states, we conclude that the event defined by
the condition Det$\rho'=0$ has zero probability and can be discarded
to compute the probabilities of $ADE (=ADE')$ or $SDE (=SDE')$. From
the form of $\rho'$ it is easy to find initial states such that
Det$\rho'$ is strictly less or strictly greater than zero, so small
balls (with positive probability) around these states also have the
same sign for this determinant. As a consequence, we have that
$P(SDE)>0, P(ADE)>0$, the actual values depend on the specific
measure used. The point is, with no additional requirement on the
measure, both situations, asymptotic or sudden death, can be found
for this dynamics.

Since this dynamics do not have asymptotic entangled states one have
$SBE=AE=\emptyset$.

\subsection*{Case 2b): More than one asymptotic states with points in the border of $S$ with $E$}

For more than one asymptotic state in this geometric situation we
can distinguish four subcases, as discussed below.

\paragraph*{All other points belong to \emph{Int}$S$.} Two non-interacting qubits subjected to two
independent phase reservoirs provide an example. The dynamics
(interaction picture implied) is given by:
\begin{subequations}
\begin{equation}\label{eqmestra}
\frac{d\rho}{dt}=\mathcal{D}_{A}\otimes I [\rho]+I\otimes
\mathcal{D}_{B} [\rho]
\end{equation}
where
\begin{equation}\label{phasedissipator}
\mathcal{D}_{i}[\rho]=\gamma(\sigma_{z,i}\rho \sigma_{z,i}-\rho),
\end{equation}
\end{subequations}
with $\gamma$ a positive constant. This dynamics may be implemented
experimentally for ions in a trap \cite{proposta}. The reservoir
would be given by applying $z$-directed magnetic fields with random
and independent magnitudes on
each ion \cite{ions} (the qubits encoded in the electronic spin of
the ions). It is easy to show that if we write the initial state in
the computational basis the evolution will be given by exponential
decays of all non-diagonal terms and all the diagonal ones will
remain constant. So the set of asymptotic states will be given by
the three real parameters set (an intersection of a four dimensional
subspace of the set of Hermitian matrices with the set of states):
\begin{equation}
\rho_{st}= \left( \begin{array}{cccc}
p_{1} & 0 & 0 & 0\\
0 & p_{2} & 0 & 0\\
0 & 0 & p_{3} & 0\\
0 & 0 & 0 & p_{4}
\end{array} \right),
\end{equation}
with $p_{i}\geq0$ for $i=1,...,4$ and $\sum_{i=1}^{4}p_{i}=1$.

In this case we have that all asymptotic states are diagonal in the
computational basis and again we have Det$\rho=$Det$\rho^{\Gamma}$.
Two situations are possible: these determinants are zero or
positive. Again, entanglement can die asymptotically or
suddenly as the following initial states illustrate:
\begin{equation}
\rho(t=0)=\left( \begin{array}{cccc}
p_{1} & 0 & 0 & 0\\
0 & p_{2} & c & 0\\
0 & c & p_{3} & 0\\
0 & 0 & 0 & p_{4}
\end{array} \right),
\end{equation}
with $|c|>0$ (as a consequence, $p_{2}>0$ and $p_{3}>0$). The
evolution will be given by states with the same form but with
$|c(t)|$ decaying exponentially, so
$d^{\Gamma}(t)=p_{2}p_{3}(p_{1}p_{4}-|c(t)|^{2})$. Then, it is
evident that, if $p_{1}$ or $p_{4}$ are initially zero, entanglement
will decay only asymptotically to a state in the border of $S$. But
if both of them are non-zero and $p_{1}p_{4}<|c(0)|^{2}$, then it
will die suddenly while the state converges to (a state in) the
interior of $S$. For $p_1p_4 \geq |c(0)|^2$ the complete trajectory
will remain in $S$.

Although examples of both situations can be provided, the typical
case is definitely sudden death of entanglement\footnote {In fact, the
condition Det$\rho$=0, implies $P(\partial D)=0$. In
this case we have also $ADE=ADE'$ and $SDE=SDE'$. Given that the
sorted state is entangled, it will exhibit SDE for sure if
Det$\rho>0$ since it will converge to a state in Int$S$. So, since
$P(\text{Int}D)=1$, we have:
$P(SDE)=P(SDE\cap\text{Int}D)=P(E\cap\text{Int}D)=P(E)$, \ie
$P(SDE|E)=1$. Equivalently, a state can exhibit ADE only if Det$\rho=0$ so
that it will converge to a state in $\partial S$, hence $P(ADE)\leq
P(\partial D)=0$.}.
As this dynamics do not exhibit asymptotic states with entanglement,
$SBE=AE=\emptyset$

\paragraph*{All other points belong to $E$.} For this case we chose a situation where both
qubits are identical (but distinguishable) and interact collectively with a common
reservoir, as it happens with two spatially close two level atoms
(close compared to the wavelength defined by their transition) in a
thermal field. The dynamics of this situation can be described by
the following master equation (also in the interaction
picture)\cite{commonreservoir}:
\begin{subequations}
\begin{eqnarray}
\frac{d\rho}{dt}&=&\gamma(2J_{-}\rho J_{+}-J_{+}J_{-}\rho-\rho
J_{+}J_{-})\nonumber\\
&&+\gamma'(2J_{+}\rho J_{-}-J_{-}J_{+}\rho-\rho
J_{-}J_{+}),\label{equacao3}
\end{eqnarray}
with $J_{\pm}=\sigma_{\pm,A}+\sigma_{\pm,B}$. A convenient way to
analyze this dynamics is to write the equations of motion for the
density matrix elements in the basis composed by the states
$\{\ket{11},\ket{\Psi_{+}},\ket{00},\ket{\Psi_{-}}\}$, resulting:
\begin{align}\label{dinresercomum}
\dot{\rho}_{11}&=-2\gamma\rho_{11}+2\gamma'\rho_{22},\nonumber\\
\dot{\rho}_{22}&=2\gamma(\rho_{11}-\rho_{22})+2\gamma'(\rho_{33}-\rho_{22}),\nonumber\\
\dot{\rho}_{33}&=2\gamma\rho_{22}-2\gamma'\rho_{33}, \nonumber\\
\dot{\rho}_{44}&=0, \nonumber\\
\dot{\rho}_{12}&=-2\gamma\rho_{12}+2\gamma'\rho_{23}-\gamma'\rho_{12},\\
\dot{\rho}_{13}&=-\gamma\rho_{13}-\gamma'\rho_{13},\nonumber\\
\dot{\rho}_{14}&=-\gamma \rho_{14},\nonumber\\
\dot{\rho}_{23}&=-\gamma\rho_{23}+2\gamma\rho_{12}-2\gamma'\rho_{23},\nonumber\\
\dot{\rho}_{24}&=-\gamma\rho_{24}-\gamma'\rho_{24},\nonumber\\
\dot{\rho}_{34}&=-\gamma'\rho_{34}.\nonumber
\end{align}
\end{subequations}

The reservoir at zero temperature corresponds to the case
$\gamma'=0$. It is easy to see from the equations of motion that
the complete subspace ${span} \left\{\ket{00},\ket{\Psi^{-}}\right\}$ is stationary
under this dynamics.
By convexity, the stationary states have the following form:
\begin{equation}
\rho_{st}=\left( \begin{array}{cccc}
0 & 0 & 0 & 0\\
0 & 0 & 0 & 0\\
0 & 0 & 1 - \rho_{44} & \rho_{34}\\
0 & 0 & \rho_{34}^{*} & \rho_{44}
\end{array} \right)
\end{equation}
and can be identified with a Bloch ball inside $D$. All states have null
determinant, so all of them are at the boundary of
$D$. It is readily seen (representing these states in the
computational basis) that Det$\rho^{\Gamma}=-(\rho_{44}/2)^{4}$, which is null
only if $\rho_{44}=0$ and is negative
 otherwise, so the set do not
have any points in the interior of $S$: this Bloch ball just touches
the set of separable states in one point. Some things can be
inferred immediately from the geometry of this set: a) the
entanglement of the system may never die (the singlet state is
stationary, for instance); b) it can be created: take any initially
separable state $\rho$ with non-zero population in the singlet
state; c) In principle, the entanglement can die asymptotically or
suddenly. In fact, initial states leading to this situation exists
but only a) and b) are ``typical''.

A helpful fact about this problem is that the singlet population is
constant through the evolution so, if this population is positive on
the initial state, it will converge to an entangled state. Since the
event formed by all states with non-zero population have probability
one, we immediately infer: $P(AE)=1, P(SBE)=P(S)$, or $P(SBE|S)=1$,
that is, if one chooses randomly an initial state, regardless if it
is entangled or not, it will evolve to an entangled state with
probability one. From this we immediately see that
$P(ADE)=P(SDE)=P(ADE')=P(SDE')=0$, i.e, the probability to choose an
initially entangled state whose entanglement will vanish is zero.

Nevertheless, one can find atypical specific examples exhibiting SDE and ADE.
Consider, for instance, the family of initial states where the only
non-vanishing matrix elements (in the basis mentioned above) are
$\rho_{11}, \rho_{22}, \rho_{33}$. From Eqs.~\eqref{dinresercomum} it
follows that those will continue to be the only non-vanishing
elements. If also $\rho_{11}=0$ their behavior is quite simple:
$\rho_{11}(t)=0, \rho_{22}(t)=\rho_{22}e^{-2\gamma t},
\rho_{33}(t)=1-\rho_{22}(t)$. So if $\rho_{22}\neq0$ the state will
remain entangled for all times (mixture of a Bell state with an
orthogonal separable state) and will die asymptotically, {\it{i.e.}},
exhibit ADE. On the other hand, if $\rho_{11}\neq 0$ the behavior of
these matrix elements is still simple and the determinant of the
partial transpose will acquire the following form:
Det$\rho^{\Gamma}(t)=\rho_{11}e^{-2\gamma t}+ P(t)e^{-4\gamma t}$,
where $P(t)$ is a second degree polynomial with coefficients
determined by the initial density matrix elements. Since
$\rho_{11}\neq 0$, this determinant will be positive after a certain
instant of time, {\it{i.e.}}, the state will be always separable after that
instant. If, {\it{e.g.}}, $\rho_{33}=0, \rho_{11}\neq0, \rho_{22}\neq0$ the
initial state is entangled and therefore will exhibit SDE.

\paragraph*{Some points belong to \emph{Int}$S$ and others to $E$.} The reservoir used in the last subcase, if taken
at positive temperature, provides this example and, to simplify the
problem we take the infinite temperature limit ($\gamma=\gamma'$ in
Eq.~\eqref{equacao3}). It is interesting that, irrespectively of
temperature, the singlet state is stationary and also the
singlet population of any state (the singlet spans a one dimensional
{\emph{decoherence free subspace}}
for this model). From the equations of motion
immediately follows that the stationary states are:
\begin{eqnarray}
\rho_{st}=\left( \begin{array}{cccc}
\frac{1-p}{3} & 0 & 0 & 0\\
0 & \frac{1-p}{3} & 0 & 0\\
0 &  & \frac{1-p}{3} & 0\\
0 & 0 & 0 & p
\end{array} \right),
\end{eqnarray}
where $p$ is the singlet population of the state. That is, they are
the Werner states (with a different parametrization).

\begin{figure}[htp]
  \centering
  \includegraphics[width=6.0 cm]{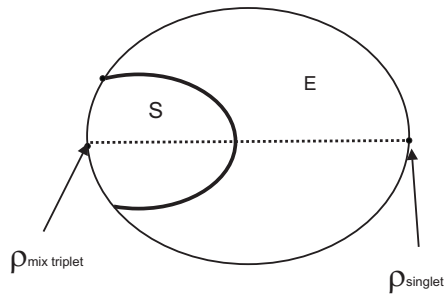}\\
  \caption{Set of asymptotic states for two qubits interacting with a common reservoir at infinite temperature. Here, $\rho_{mix~triplet}=\frac{1}{3}(\ket{11}\bra{11}+\ket{\Psi_{+}}\bra{\Psi_{+}}+\ket{00}\bra{00})$.}
  \label{figura2}
\end{figure}

The determinant of the partial transpose (with respect to the computational
basis, of course) is simply $(3-12p^{2})/36$ being negative only if
$p>1/2$. The set of stationary states forms
a line segment in $D$ with both ends, those with $p=0$ or $p=1$, on
the border of $D$, one of them in the interior of $S$ (relative to
$D$) and the other in $E$, respectively, and the line intersecting
the border between $S$ and $D$ when $p=1/2$ (see
Fig.~[\ref{figura2}]).

Since the singlet population remains fixed in the
dynamics it allows us to identify the asymptotic state of any given
initial condition. An initial state will have non-zero entanglement
asymptotically if, and only if, $\rho_{44}>1/2$ so we have
$P(AE)=P(D_{>1/2}=\{\rho\in D|\rho_{44}>1/2\})>0$. Of course
$P(SBE)=P(D_{>1/2}\cap S)$. Since a state can exhibit ADE iff it
relaxes to a state in the border between $S$ and $E$, we
have $P(ADE')\leq P(ADE)\leq P(ADE\cap\{\rho\in
D|\rho_{44}=1/2\})=0$. So ADE is atypical for this dynamics but SDE,
on the other hand, have a non-zero probability. In fact, an
initially entangled state have SDE iff $\rho_{44}<1/2$,
so $P(SDE')=P(E\cap \{\rho\in D|\rho_{44}<1/2\})>0$.

\paragraph*{All points belong to $\partial S$.} The combination of two reservoirs used in former examples will
provide this case. If we have qubit $A$ subjected to spontaneous
decay and $B$ to a phase reservoir the system will have the desired
behavior, a situation that may occur experimentally if we entangle
an atom in vacuum with a spin subjected to a stochastic magnetic
field. That is, the system dynamics would be described by a master
equation of the form \eqref{eqmestra} (again in the interaction
picture), but with $\mathcal{D}_{A}$ given by
Eq.~\eqref{decaydissipator} (with $i=A$ and $\gamma_{A}'=0$) and
$\mathcal{D}_{B}$ by Eq.~\eqref{phasedissipator} (with $i=B$). It is
easy to see that the set of asymptotic states will be constituted by
the product states where $A$ is in the $\ket{0}$ state and $B$ in a
state described by a diagonal matrix (in the computational basis),
so the global states reads:
\begin{eqnarray}
\rho_{st}=\left( \begin{array}{cccc}
0 & 0 & 0 & 0\\
0 & p & 0 & 0\\
0 & 0 & 0 & 0\\
0 & 0 & 0 & 1-p
\end{array} \right),
\end{eqnarray}
for $0\leq p\leq1$. Whatever the value of $p$ we have
Det$\rho_{st}$=Det$\rho_{st}^{\Gamma}=0$ so they indeed belong to
$\partial S$. Again we have $SBE=AE=\emptyset$ for this dynamics,
since there are no entangled asymptotic states, but to analise the
probability of the other events we use the exact solution for the
dynamics and write the determinant of the partial transpose in the
form:
\begin{equation}
\text{Det}\rho(t)^{\Gamma}=f_{1}(\rho)e^{-\lambda_{1}t}+...+f_{n}(\rho)e^{-\lambda_{n}t},
\end{equation}
where the functions $f_{i}$ depend on the initial state only, while
$\lambda_{1}<\lambda_{2}<...<\lambda_{n}$. In this way, as long as
$f_{1}\neq0$, the asymptotic sign of Det$\rho(t)^{\Gamma}$ will by
given by the sign of $f_{1}$. Denoting by $\gamma_{A}$ and
$\gamma_{B}$ the decay rate for each reservoir, it so happens that
$\lambda_{1}=2 \gamma_{A}$ and $f_{1}=\text{Det}\rho'$, where:
\begin{eqnarray}
\rho'=\left( \begin{array}{cccc}
\rho_{11} & \rho_{12}^{*} & 0 & 0\\
\rho_{12} & \rho_{11}+\rho_{22} & 0 & 0\\
0 & 0 & \rho_{33} & \rho_{34^{*}}\\
0 & 0 & \rho_{34} & \rho_{33}+\rho_{44}
\end{array} \right).
\end{eqnarray}
Since this matrix is positive definite (given that $\rho$ is), the
system will reach its asymptotic state from the interior of the
separables if $f_{1}(\rho')>0$. But the event $f_{1}(\rho')=0$ have
zero probability, so we may conclude that $P(SDE)=P(SDE')=P(E)$,
while $P(ADE)=P(ADE')=0$, that is, a sorted initial entangled state
will exhibit sudden death of entanglement with certainty, in contrast
with case 2a) where sudden and asymptotic death both had positive
probabilities. Still, it is possible to find specific states where
asymptotic death takes place. For instance, consider the set of
initial states:
\begin{eqnarray}
\rho=\left( \begin{array}{cccc}
0 & 0 & 0 & 0\\
0 & \rho_{22} & \rho_{23} & 0\\
0 & \rho_{23}^{*} & \rho_{33} & 0\\
0 & 0 & 0 & 0
\end{array} \right).
\end{eqnarray}
The partial transpose determinant will be then
Det$\rho(t)^{\Gamma}=-|\rho_{23}(t)|^{2}\rho_{22}(t)\rho_{33}(t)$,
being negative for all $t$ if $\rho_{23},\rho_{22}$ and $\rho_{33}$
are initially different from zero, so the entanglement dies
asymptotically.

\subsection*{Case 3a): One asymptotic state in $E$}

The most natural way to realize a dynamics with this property is
through a thermal reservoir. This time, though, interacting qubits
and a common thermal reservoir are needed, that is, a reservoir that
take any initial state to the Gibbs state $Z^{-1}\exp({-\beta
H})$, with $Z=\text{Tr}\exp({-\beta H})$ and $H$ stands for the Hamiltonian
describing the closed dynamics of the qubits. Typically, the ground
state of interacting qubits Hamiltonian is non-degenerate and
entangled, so if $\beta$ is large enough we obtain the desired dynamics.

Dynamics with these asymptotic states can be engineered using Lindblad
autonomous equations, at least formally. Actually, fixed an
arbitrary state for the system, there are many Linbdbladians that
have this state as the only asymptotic state, in particular there
are ones with only one Lindblad operator and null Hamiltonian part
\cite{dietz}. The specific Lindbladian of course, will depend on the
specific interaction between qubits and reservoir (if the dynamics
could be described by a Lindblad equation in the first place).

To give a more specific picture, consider, for instance, two
interacting qubits described by the following Hamiltonian:
\begin{equation}
H=\frac{1}{2}\omega\sigma_{z,A}+\frac{1}{2}\omega\sigma_{z,B}+g(\sigma_{+,A}\sigma_{-,B}+\sigma_{-,A}\sigma_{+,B})
\end{equation}
with $\omega, g$ positive constants satisfying $g>\omega$ (\ie strong coupling limit). The
eigenvalues for this Hamiltonian are, in crescent order,
$-g,-\omega,\omega,g$, with respective eigenvectors
$\ket{\Psi_{-}},\ket{00},\ket{11},\ket{\Psi_{+}}$,
leading to an entangled ground state. Denote by $\ket{i}$,
$i=1,\ldots,4$ these eigenvectors according to their eigenvalues
order. We may consider a thermal reservoir at null temperature that
induces decays between any two of these states in a Markovian way,
such that the dissipator would be:
\begin{equation}
\mathcal{D}[\rho]=\sum_{i<j}\gamma_{ij}(2\sigma_{ij}\rho\sigma_{ji}-\sigma_{jj}\rho-\rho\sigma_{jj}),
\end{equation}
where $\sigma_{ij}=\ket{i}\bra{j}$ and $\gamma_{ij}$ are
non-negative constants. A dissipator of this type can be derived
from a microscopic model, for instance, adapting the results of Ref.
\cite{scala} to the Hamiltonian considered here .

As in case 1a), the events and probabilities we are interested in
are trivial: $SBE=S, AE=D$, {\it{i.e.}}, every initial
state will acquire entanglement for large times, in particular the
separable ones, so $P(SBE)=P(S)$ and $P(AE)=1$. Since the
entanglement never vanishes, $ADE=ADE'=SDE=SDE'=\emptyset$

\subsection*{Case 3b): Several asymptotic states in $E$}

Examples for this case can be provided just by the same trick used
in case 1b): we take any Lindbladian with only one asymptotic
entangled state and insert a time varying coupling which multiplies
the dissipator. The same reasoning can be applied with respect to the
asymptotic states for the subsequent dynamics (in the interaction
picture), so, if the decay rate of the coupling is small enough, the
set of asymptotic states will be constituted by a small ``blurring''
around the asymptotic state of the dynamics with constant coupling.

Contrary to case 1b), though, in what entanglement is concerned, it
is important now whether the dynamics is given in the Schr\"{o}dinger or
interaction pictures, because their correspondence is given by
global unitary transformations. By the same reason as before, the
dynamics will not be relaxing in the Schr\"{o}dinger picture, but
one can still find a non-trivial asymptotic set, this time, entirely
contained in $E$. Indeed, diminishing the decay rate of the
reservoir couplings, we can diminish at will the diameter of the set
of stationary states in the interaction picture which, by its turn,
always contain the Gibbs state of the system. Now, unitary
transformations are isometries for practically all relevant metrics,
so the set of asymptotic states in the interaction picture is mapped
to sets with the same diameter in the Schr\"{o}dinger picture. But
these unitary transformations have the Gibbs state as a fixed
point, hence these sets always contains it. Since $E$ is open, given
that their diameter is small enough, we can be sure that they always
fall entirely inside of it.

As a consequence of the discussion in the above paragraph, the
events and probabilities we are considering in this paper are
identical to the ones in the last example.

\section{Conclusions}

In this paper we review the classification of the possible dynamics of entanglement
based on the relative geometry of the sets of asymptotic and separable states.
We provided
examples for all possible classes, including the previously unknown cases with more than one
asymptotic state, but avoiding the boundary $\partial S$. In giving those
examples it was sufficient to use
two-qubit dynamics dictated by equations of motion in
Lindblad form (including non-autonomous dynamics exactly for those previously
hard examples). In each case, the
existence of sudden death of entanglement, asymptotic death of
entanglement, sudden birth of entanglement and asymptotic
entanglement were analyzed from a more precise point of view,
looking at the \emph{probabilities} that each of these phenomena
occur if one choose a random initial state and a suitable
probability measure on the set of quantum states.

\begin{acknowledgments}
We thank CNPq and FAPEMIG for financial support. This work is part
of the Brazilian National Institute for Science and Technology on Quantum Information.
\end{acknowledgments}

\end{document}